\documentstyle[11pt]{article}
\newcommand{\blankline}{\vskip .3cm}
\newcommand{\f}{\begin{equation}}
\newcommand{\ff}{\end{equation}}
\begin{document}
\rightline{\Large CGPG-95/8-7}
\centerline{\LARGE  The Bekenstein bound,}
\blankline
\centerline{\LARGE   topological quantum field theory}
\blankline
\centerline{\LARGE and pluralistic quantum cosmology}
\rm
\vskip.3cm
\centerline{Lee Smolin${}^*$ }
\blankline
\centerline{\it  Center for Gravitational Physics and Geometry}
\centerline{\it Department of Physics, The Pennsylvania State
University}
\centerline{\it University Park, PA, USA 16802}
\blankline
\centerline{\today  }
\centerline{ABSTRACT}
An approach to quantum gravity and cosmology is proposed based
on a synthesis of four elements: 1) the Bekenstein bound and the
related holographic hypothesis of 't Hooft and Susskind, 2) topological
quantum field theory, 3) a new approach to the interpretational
issues of quantum cosmology and 4)  the loop representation
formulation of non-perturbative quantum gravity.  A set of
postulates are described, which define a {\it pluralistic quantum
cosmological theory.} These incorporates a statistical and relational
approach to the interpretation problem, following proposals of Crane
and Rovelli, in which there is a Hilbert space associated to each
timelike boundary, dividing the universe into two parts.  A quantum
state of the  universe is an assignment of a statistical state into each
of these  Hilbert  spaces,  subject to certain conditions of consistency
which come from  an analysis of the measurement problem.   A
proposal for a concrete realization of these postulates is described,
which is based on certain results in the loop representation and
topological quantum field theory, and in particular on the
fact that spin networks and punctured surfaces appear in both
contexts.  The Capovilla-Dell-Jacobson solution of the constraints of
quantum gravity are expressed quantum mechanically in the language of
Chern-Simons theory, in a way that leads also to the satisfaction of
the Bekenstein bound.

\noindent
\blankline
\eject
\section{Introduction}

In this paper a new approach to the problem of constructing a
quantum theory of gravity in the cosmological context is proposed.
It is founded on results from four separate directions of
investigation,
which are:

1)  A new point of view towards
the interpretation problem in quantum
cosmology\cite{louis,carloqm,forabner,lotc},  which
rejects the idea that a single quantum state, or a single Hilbert
space, can provide a complete description of a closed system like
the universe.  Instead, the idea is to accept Bohr's original
proposal that the quantum state requires for its interpretation
a context in which we distinguish two subsystems
of the universe -the quantum system
and observer.  However, we seek to relativize
this split, so that the boundary between the
part of the universe that is considered the system and that which
might be considered the observer may be chosen arbitrarily.
The idea is then that a quantum theory of cosmology is specified
by giving an assignment of a Hilbert space and algebra of
observables to every possible boundary that can be considered to
split the universe into two such subsystems.  A quantum
state of the universe is then an assignment of a statistical
state to every one of these Hilbert spaces, subject to certain
conditions of consistency.   Each of these states is interpreted to
contain the information that an observer on one side of each
boundary might have about the system of the other side.

This formulation then accepts the idea that each observer
can only have incomplete information about the universe,
so that the most complete description possible  of the universe
is given by the whole collection of incomplete, but mutually
compatible quantum state descriptions of all the possible
observers.  At the same time, the information of different
observers is, to some extent, different, so that there is no way,
in principle, to combine the descriptions of each observer to
construct a single quantum state that could give a complete
description of the whole universe.
This point of view, which has been developed in collaboration
with Louis Crane\cite{louis} and Carlo Rovelli\cite{carloqm} and
is discussed also
in papers by them,  may
be called pluralistic quantum mechanics.

2)  The Bekenstein bound\cite{bekenstein}, which requires, for
reasons
that will be reviewed in section 2, below, that the Hilbert
spaces associated
to  timelike
boundaries of fixed spatial area
$A$ must have finite dimension proportional to
$exp(c A/l_{Planck}^2)$, with $c$ some fixed dimensionless
constant.    This has recently led 'tHooft\cite{gerard-holographic}
and Susskind\cite{lenny-lorentz,lenny-bh} to
make the ``holographic hypothesis", according to which quantum
gravity must be understood to be constructed from field theories
on two dimensional surfaces, that describe what knowledge
an observer looking through the surface at the world on the other
side, might have.  The question has then been raised as to
what specifies the surfaces on which these ``holographic
field theories" are to be defined.  The answer pluralistic
quantum theory gives to this question is that the ``holographic"
description must be applicable to any possible surface, so that the
task of the theory is not to pick out certain surfaces but to provide
the conditions and relationships that hold between the theories
on the different surfaces.

3)  Topological quantum field theory\cite{witten-tqft,atiyah,segal}
which, first of all, provides
examples of pluralistic quantum theories\cite{louis}, and, secondly,
gives us a set of finite dimensional
Hilbert spaces with which to realize the Bekenstein bound.

4)  The loop representation approach to non-perturbative quantum
gravity \cite{lp1,lp2,carlo-review,ls-review,aa-review}, and,
in particular, the kinematical part of the theory,
associated with the description of spatially diffeomorphism invariant
states and observables in terms of spin networks
\cite{ls-review,weave,volume,renata,spinnet-us,ham1,spinnet-john,gangof5}.
In particular,
we may make use of the fact that the same mathematical structure-
spin networks\cite{roger-spinnet}- appears
as labels of diffeomorphism invariant states
in both the loop representation formulation of quantum gravity
and in topological quantum field
theory\cite{witten-tqft,qnet,louis2d3d}.

One of the motivations for the present work is a study
recently carried out by the author of
quantum general relativity in the presence of a particular set of
boundary conditions\cite{linking}.  In this context,
the appearance of
the spin networks as labels of states in these two theories
was used to construct
state spaces that represent the algebra of observables of general
relativity on a timelike boundary out of the Hilbert spaces of
certain Chern-Simons theories.  It was then natural to contemplate
the conjecture that the observables measured on the boundary
are sufficient to determine the quantum state of the system.
If this conjecture is true it means
that the whole physical state space for quantum general
relativity, in the presence of these boundary conditions,
can be constructed
from direct products of state spaces for Chern-Simons theory.
Further, the correspondence
between spin and area leads to the result that the state spaces
for quantum gravity constructed in this way satisfy the Bekenstein
bound.  This is because they have the property that
once the metric of the boundary is measured, the
space of states accessible by measurements on the boundary is
reduced to a finite dimensional subspace, whose dimension
grew with the exponential of the area.

This work was carried out in a less general context than that
contemplated here, in which state spaces are to be associated with
any possible timelike boundary dividing the universe into two
parts, no matter what conditions are satisfied on it.  However, its
results are part of the motivation for this proposal.   Indeed, I will
argue below that general features of this construction, and in
particular the decomposition of state spaces for quantum gravity
in terms of the Hilbert spaces of topological quantum field theories,
may be realized in this more general context.

Another motivation for this proposal is to
try to reverse the usual strategy of constructing a quantum theory
of gravity and to take the point of view that, rather than
deriving that theory directly from the classical theory by following
some more or less well defined quantization procedure, the
goal is to formulate a set of postulates that define
the theory directly\cite{exp}.  The inspiration for these postulates is certain
results and conjectures which have arisen in the course of
investigations of the quantization of classical gravitational
theories.   The
hope is that there may be a set of such postulates that are sufficient to
found a theory which is well defined, has a meaningful physical
interpretation and at the same time allows both classical general
relativity and quantum field theory on fixed backgrounds
emerge as approximations in particular
limits\footnote{An important motivation for this
idea is the recent work of Jacobson\cite{ted-new}, where he does
exactly
this, by deriving classical general relativity from the connection
between area and information (essentially the Bekenstein
bound) and the laws of thermodynamics.}.

The reason for taking such an approach lies in the peculiar situation
that research in quantum gravity has led to, in which
several different
approaches have yielded results that we may hope will be
preserved in the final theory, while at the same time, no single
approach has led to the formulation of a complete theory.
Furthermore, there are, in each case, reasons to believe that
the successful results of each approach may not easily be
recovered in the others approaches.  It may then be that what we
have in each case is a partial theory that describes successfully
some domain of quantum gravitational phenomena, but which does not
tell the whole story.

Here, I am referring mainly to three approaches: string theory,
quantum field theory in curved space-time and canonical
quantization of general relativity based on the merging of the
Ashtekar variables with the loop representation.  For
example the latter approach has yielded a kinematical
description of quantum general relativity in terms of
(three dimensional) diffeomorphism
invariant states and observables
that has many appealing features.  These include the
prediction of discrete spectra for areas and
volumes\cite{ls-review,weave,volume,renata,spinnet-us,gangof5}
the description of states in terms of spin networks,  and the
suggestions of the existence of a linearized sector with a
natural Planck scale cutoff\cite{carlojuniche}.
Some of these results even turn out
to be derivable in a completely rigorous
mathematical framework\cite{gangof5}.
At the same time, despite very exciting progress in the last
year, this approach has yet to completely  resolve the difficulties
concerning physical observables, time and the inner product which
plague it and, indeed, all canonical approaches to quantum
gravity\cite{timeproblem}.
As a result
of these difficulties, it is still not known whether the exact
physical states\cite{tedandi,lp1,jorgerodolfo}, which were
among the
first results of this approach,
are physically meaningful or useful.

This does not mean that
this approach is not useful.  It may lead to a completely well
defined mathematical formalism which, however, suffers difficulties
of interpretation coming from the problems of extending quantum
theory to the cosmological case. It may
be possible in certain contexts to overcome some of these
difficulties by defining approximation procedures
which yield physically meaningful predictions in restricted
circumstances\cite{deiter,ham1,carlot,ham2}.  Still,
unless the much discussed fundamental difficulties facing
such theories\cite{timeproblem} are overcome, we still may not
have a satisfactory quantum theory of gravity.

Another crucial issue for this kind of approach is the recovery of
Lorentz invariance in the appropriate limit.
In fact, the restoration of Lorentz invariance is an issue
for any quantum field theory one of whose parameters is
a length which serves
as an invariant
cutoff, for the following reason.  An acceptable quantum theory of
gravity must have a limit in which the physics can be
described in terms of small excitations of a Lorentz invariant
vacuum.
But, the existence of a cutoff scale means that if an inertial
observer, in the presence of this vacuum, measures the
spectrum of graviton states he should see
no gravitons  with wavelengths
smaller than $l_{cutoff}$.  However, if the vacuum
is Lorentz invariant, a second inertial observer,
moving with a large $\gamma$ with respect to the first,
should see gravitons with all wavelengths longer than
$l_{cutoff}$ in their frame.  But this means that they will see
gravitons that have wavelength $\gamma^{-1} l_{cutoff}$
in the first observers frame.  We have an apparent contradiction.

Another way to see this problem is to ask whether Lorentz
invariance can be consistent with the requirement that the Hilbert
space describing the physics in any region of finite volume
must be finite dimensional, as is required by the Bekenstein
bound.  At first sight it seems that the
answer must be no.  In conventional quantum field theory it
is possible to construct wavepackets that describe quanta
moving with any peak wavelength $\lambda_0$, with
a spread $\delta \lambda_0 < L$ where $L >> \lambda_0$
is the linear
size of the region.  Thus, the existence of finite volume
boundary conditions
does not prevent the theory from having states of arbitrarily
small wavelength.  And, in a linear field theory, all these
states are orthogonal to each other, which means that there
are an infinite number of orthonormal states.  Thus, it seems
that Lorentz invariance cannot be consistent with a
theory that has a finite number of degrees of freedom per
fixed spatial region.

It is then very impressive that there is one context in which
this problem has been definitely
solved, which is perturbative string
theory\cite{strings,he,attickwitten,stringdiscrete,lenny-lorentz,ks}.
The problem is solved there because the elementary excitations
are extended one dimensional objects.  As is explained in detail
in \cite{lenny-lorentz,ks}, string theory is consistent
with Lorentz invariance in
spite of having a finite number of degrees of freedom per fixed
spatial
region because the strings, representing the small excitations
of the vacuum, can diffuse transversally as they are boosted
longitudinally.

A theory of extended objects is different in this respect
from a theory of
pointlike objects, because any extended excitation of
a theory that is both Lorentz invariant
and finite must, if boosted sufficiently
become effectively two dimensional, as its longitudinal extension
is contracted below the scale of the cutoff.  This means that
extreme relativistic excitations may be naturally described as
excitations of a two dimensional field theory.
It is then very interesting that, as was shown by Klebanov
and Susskind\cite{ks}, continuum string theory can emerge from a
lattice field theory in which there is a  cutoff in the
transverse directions by means of
a limit in which the lengths of the
strings diverge while the transverse cutoff remains fixed.

Given that string theory solves this problem and, more generally,
that the only known perturbative quantum field theories that
describe consistently the coupling of gravitons to themselves
and other fields are perturbative string
theories\cite{strings}, it seems that
any acceptable quantum theory of gravity, whatever its
ultimate formulation, is likely to reduce to a perturbative
string theory in the appropriate limit.  It is, of course, possible
to imagine that non-perturbative quantum general relativity
will, for this reason, have some perturbative string theory as an
appropriate limit\cite{exp}. However, while this is a possibility
that must be explored, it's success is certainly not guaranteed by
what we know about the theory so far.

At the same time, string theory cannot be itself the whole theory
unless it has a nonperturbative formulation.  While there have
been recently some very interesting hints, coming from the
very significant discovery of non-perturbative effects in string
theory\cite{newstringstuff},
it is still unclear to what extent string theory will have
to be modified to arrive at a completely non-perturbative
formulation of the theory.  Thus, counting also quantum field
theory in curved space-time, our situation is that we have
interesting results, and in some cases even physical
predictions\cite{exp}, coming from three different starting
points, each of which, however, may not lead to a complete
theory.  In this circumstance it may be prudent to try
to find a new starting point that combines what is useful
about each of these directions.

Another way to see that a new starting point may be needed is
that exactly to the extent that these three well explored approaches
are successful, they cast doubt on the physical
relevance of their starting points, which in each
case involve the quantization
of conventional classical field theories.
The
results of each of these three directions seem indeed to
indicate that there is a
natural Planck scale cutoff, below which physical degrees
of freedom cannot be resolved.
This suggests that, ultimately, a quantum theory of gravity
will not be formulated most simply as a theory of fields on
a differential manifold representing the idealized-and
apparently nonexistent- ``points" of space
and time\footnote{Of course, that space and/or time
must be fundamentally discrete is an old idea,
for a good review, see \cite{discrete}}.

To put this another way, the space of fields-the basic
configuration space of classical field theory-has
been replaced in the quantum theory by abstract Hilbert
spaces.   At the same time, ordinary space, in these formulations,
remains classical, as it remains the label space for the field
observables.  This perpetuates the idealization of arbitrarily
resolvable space-time points, that the results of string theory,
non-perturbative quantum gravity
and semiclassical quantum gravity (through the
Bekenstein bound) suggest we must give up.

The final motivation for the present approach comes from the
interpretational problem in quantum cosmology.  Despite very
interesting proposals
\cite{everett,dwg,griffiths,omnes,gellmann-hartle,hartle,zeh,zurek,simon}
it may be said that
no proposal to interpret the conventional Hilbert space formulation
of quantum cosmology has so far
convincingly succeeded\cite{dowkerkent}. (I will
discuss this issue in some detail in section 4, below).  This means
that even if the problem of physical observables could be solved
in the context of conventional Hamiltonian quantization, perhaps as
envisioned by Rovelli\cite{carlo-time} or recently by
Ashtekar, Lewandowski,
Marlof, Mour\~{a} and Thiemann\cite{gangof5}, the
resulting mathematical
theory would still not have an acceptable physical interpretation.
More precisely, as I will argue, while the theory might have
a fanciful interpretation in terms of the observations made by
some ``God" outside of the universe, this would be of no help
to connect the mathematical framework to
observations made by us observers who live in the universe.
For this reason also, it may be useful to contemplate
taking a new point of view according to which quantum cosmology
might have an interpretation strictly in terms of information that
might be held by observers inside the universe.  This, as I will
describe, is another goal of the present proposal.

This paper is divided roughly into two parts, the first
of which is more general and motivational, while the
second is more focused and mathematical. The
next three sections motivate the idea of a pluralistic
formulation of quantum cosmology.  Sections 2 and 3
describe the argument for, and the implications
of, the Bekenstein  bound.  Section 4 is devoted to the
problems of the interpretation of quantum cosmology,
and give an introduction to the main ideas of the approach
of Crane\cite{louis}, Rovelli\cite{carloqm} and
the author\cite{forabner,lotc} for a pluralistic
approach to quantum cosmology.  The postulates of the
theory are then presented at the end of section 5.

 Section 6 is devoted to the problem of time. I show that
this problem can be circumvented in pluralistic quantum cosmology
by, as it were, standing on their heads certain proposals concerning
the problem of time which are usually set in the conventional
formulations
in which  a single
state and Hilbert space describe the universe.  These proposals,
which fail in one way or another in that context, take on a
somewhat different aspect in the context of pluralistic quantum
cosmology.  The problem of
the classical limit, and the related question of the timelike initial
date problem are the subject of section 7.

Sections 8 and 9 are more mathematical, and concern one attempt
to realize the postulates of pluralistic quantum cosmology
by building a theory on certain results of non-perturbative
quantum gravity and topological quantum field theory.
A representation for the connection and frame fields of the
Ashtekar formalism of general relativity, pulled back into an
arbitrary surface, is given in which the state space is constructed
from certain direct sums of representation spaces of quantum
Chern-Simons theory.  This formalism incorporates a proposal for
solving the constraints of the theory, in the form given by
Reisenberger\cite{mikerconstraints}, quantum mechanically.
One consequence of this
formalism is that the Bekenstein bound is automatically satisfied.

The paper closes with a short summary of conclusions and open problems.

\section{The Bekenstein argument}

 Consider a region of space, $\Sigma$,
the quantum dynamics of which we wish to study.  We will
assume that the restriction of the system of interest to
$\Sigma$ is enforced by certain boundary conditions,
defined as conditions the fields must satisfy on the spatial
boundary of $\Sigma$, which we will denote $\cal S$.
Given these, we will assume we can
define a quantum theory which is described by
a Hilbert space ${\cal H}_\Sigma$, and an algebra of
physical observables ${\cal A}_\Sigma$.  Among these
there is an important subalgebra, ${\cal A}_{boundary}$,
which consists of those physical observables which are
functionals only of fields on the boundary.  Among these
we will assume are
the  Hamiltonian, $H_\Sigma$ and the areas of regions
$\cal R$ of
the boundary $\cal S$, which I will denote $A[{\cal R}]$.
Recall that in general relativity the Hamiltonian
is, up to terms proportional to constraints,
defined as an integral on the boundary and is thus
an element of ${\cal A}_{boundary}$.

Since the system contains gravitation, we may assume that
among the spectrum of states are a set which correspond to
black holes.  These are semiclassical statistical states, and
we will assume that their real statistical entropies
are given by the usual formulas, at least
in the semiclassical limit when their masses and areas are
large in Planck units.

The argument is simplest in the case that we assume that
the induced metric in $\cal S$ is the two sphere metric.
It proceeds by assuming that the region
$\Sigma$ can  contain an object $\cal O$ whose complete
specification requires an amount of information
$I_{\cal O}$ which is larger
than
\f
I_{\cal S}= {A[{\cal S} ] \over 4 l_{Pl}^2}
\ff
which is of course the entropy of a black hole whose horizon just
fits inside of $\cal S$.

Let us assume that initially we know nothing about
$\cal O$, so that
$I_{\cal O}$ is a
measure of the entropy of the system.
However, with no other information we can conclude
that $\cal O$ is not a black hole,
as the largest information that could be  contained in any black
hole in $\Sigma$ is $I_{\cal S}$.  We may then argue, using the
Hoop theorem\cite{hoop} that the energy
contained within $\Sigma$ (as measured either by a quasilocal
energy on the surface or at infinity)
must be less than that in a black hole
whose horizon has area ${\cal A}[{\cal S}]$.   But this being the case
we can now add energy to the system to bring it up to the mass
of that black hole, which has the result of transforming $\cal O$ into
the black hole whose horizon just fits inside the sphere $\cal S$.

This can be done by dropping quanta slowly into the black hole,
in a way that does not raise the entropy of its exterior.  As a
result, once the black hole has formed we
know the entropy of the system, it is
$I_{\cal S}$.  But we started with a system with
entropy $I_{\cal O}$, which we assumed is larger.  Thus, we have
violated the second law of thermodynamics.  The only way
to avoid this is if $I_{\cal O} <  I_{\cal S}$.

We may remark that this argument employs a mixture of
classical, statistical and semiclassical reasoning.  For example,
it assumes both that the hoop theorem from classical general
relativity applies, at least in the case of black hole masses
large in Planck units, to real, quantum black holes.  One might
attempt to make a detailed argument that this must be the
case if the quantum theory is to have a good classical limit.
However worthy of a task, this will not be pursued here,
as it is unlikely that any such argument can be elevated to
establish the necessity, rather than plausibility of the
Bekenstein bound, in the absence
of a complete theory of quantum gravity.  Perhaps it is
sufficient then just to note that in twenty years no plausible
counter argument to the Bekenstein bound has survived.

\section{Consequences of Bekenstein's bound}

The Bekenstein bound has profound consequences for
the question of how a quantum theory of gravity might be
formulated.  If the entropy of
a quantum system is bounded above by $S_{max}$,
it means that that system
must have a finite dimensional state space, whose dimension
is given by
\f
dim {\cal H}=e^{S_{max} Ln(2) } =e^{c A[{\cal S}] \over l_{Pl}^2}
\ff
where $c$ is a fixed dimensionless constant.
This is true because a system whose entropy has an absolute
upper limit can, in principle, be completely specified by giving the
answers to a finite number of yes-no questions.
Thus, not only can a quantum theory of gravity not
be a conventional
quantum field theory, with an infinite number of degrees of
freedom per finite amount of spatial volume, it cannot even be
a conventional cutoff quantum field theory with a fixed finite
number of degrees of freedom per spatial volume.  Instead,
we have a fixed number of degrees of freedom per unit area
of the surface of the region.

This has led 't Hooft\cite{gerard-holographic} and
Susskind\cite{lenny-lorentz,lenny-bh} to make the
{\it holographic hypothesis},
which is that a theory of quantum gravity
which describes a region bounded by a spatial surface $\cal S$
may be described by a quantum theory with a finite number
of degrees of freedom on the surface $\cal S$.

 There are, however, two difficulties with the holographic hypothesis
which must be faced.  The first is what kind of quantum theory
will live on the surfaces.  It seems challenging to imagine
a kind of a field theory that could both provide a realistic
description of interacting gravitational and matter fields, but
is at the same time based on  finite dimensional
Hilbert spaces. Even the harmonic oscillator has
an infinite dimensional Hilbert space.    't Hooft has explored the
possibility that such a theory might be constructed from
cellular automata.  While this is an intriguing suggestion, in this
paper a different proposal will be explored, which is that these
theories are constructed, in a certain way to be described,
from topological quantum field theories.

The second question that must be confronted
is to what two dimensional
surfaces are we to associate the hypothetical holographic
field theories.  Closely related to this is another worry.
Most of the arguments given for
the holographic hypothesis involve static situations.  As a result, it is
not completely clear whether the boundaries on which
the field theories are to be defined should be in the general
case two dimensional or three dimensional.

To address these issues, we must
leave to one side for a moment  the arguments of
Bekenstein, 't Hooft and Susskind and consider a different
set of problems, that seem at first sight to be completely
independent.  These are the interpretational problems of
quantum cosmology.

\section{Pluralistic quantum cosmology}

In this section I motivate the main conceptual parts of the
proposal of this paper.  I begin with a review of the
interpretational problems
that the new proposal is intended to address.

\subsection{Summary of the problem of the interpretation of
quantum cosmology}

The interpretational difficulties with
quantum cosmology arise
because the conventional interpretations of quantum theory
require that the quantum state description be applied only
to subsystems of the universe.  The interpretation of the
theory requires the existence of things which are in the
universe but outside of the system described by the quantum
state, including the measuring instruments, the clocks that
give meaning to the Schrodinger evolution and the observers.

There is, of course, a long history of attempts to modify the
interpretation of the theory, while keeping the formalism
fixed, in order to overcome the dependence of the quantum
description on a split of the world into two parts, so that
quantum theory may be applied to the universe as a whole.
If I may briefly summarize the history of this story, it
began with the many worlds interpretation of
Everett\cite{everett},
which may be put in several different forms, some more
metaphysical\cite{dwg} and some more operational\cite{me-many}.
However, all forms
of the many worlds interpretation suffer from an ambiguity,
which is called the preferred basis
problem\cite{zurek,me-many}.  The
connection
of the theory to actual observations requires the selection
of a preferred basis, corresponding to observables we use to
describe the real world.  The theory itself seems to provide
no such choice.  A number of interesting ideas have been
tried to solve this problem\cite{zurek,zeh},
but none were completely successful.

The last few years attention shifted towards one kind of proposal
to solve this problem,
which is the consistent or decoherent
histories program\cite{griffiths,omnes,gellmann-hartle,hartle}.
The basic ideas here are two: first to interpret
the theory in terms of sequences of projection operators, which
pick out a history of quantum mechanical observations, and
second, to give an interpretation only to measurements of
consistent or decoherent sets of such histories, which are those
for which the interference effects vanish, either absolutely or
``for all practical purposes".  This program was based on the
observation that sets of alternative
histories that correspond to classical
motions do have the property of being consistent, at least to
an enormously good approximation.

However, this program is also facing a difficulty analogous
to the preferred basis problem because, as shown in a recent
paper by Dowker and Kent\cite{dowkerkent},  there are an
infinite number
of equally consistent sets of histories, most of which
 do not correspond to anything that could
be described in classical language.    Thus, while it is
true that there are sets of consistent histories that describe
evolution in terms of quasi-classical variables, these are not
distinguished by the formalism of the quantum theory.  Thus,
a selection principle, external to the formalism of the theory,
seems to be needed to make a connection between the theory
and what is actually observed.

Very interesting ideas which relate this selection problem to the
existence of self-organized complexity have
been proposed\cite{gellmann-hartle,simon}, and
are very much worth exploring.  However,
in light of this situation it may be
interesting to examine other approaches, that require the
formalism of quantum theory to be modified in order to
extend it to a theory of the whole universe.

\subsection{The basic idea of pluralistic quantum cosmology}

I would like to describe here an approach to such a modification,
that has evolved over a number of years through conversations
with Louis Crane and Carlo Rovelli.

I would like to introduce this by stepping back for a moment and
asking what the context of the problem is.  The problem of extending
our physics from a description of isolated systems to a description
of the universe as a whole involves us in a number of profound
issues, one of which is the question of what an observable
in such a theory might be.  Both abstract
argument\cite{leibniz,mach,pierce,me-ste}, and the
example of our one successful cosmological theory, which is
general relativity\cite{john-hole,julian-gr}, tell us that the
notion of what is observable
must be different in certain aspects in a cosmological theory
then it is in a theory of an isolated system.  Indeed, general relativity
differs from Newtonian physics profoundly in its formulation of
observables, because it is based on a relational view which denies
the existence of any nondynamical background structure that
may give
meaning to observables, such as coordinates or an a priori notion
of time or spatial geometry.  Instead, all observable quantities
are defined in terms of relationships between dynamical
degrees of freedom.
This relational conception of observables,
which has its origins in Leibniz's \cite{leibniz} and Mach's \cite{mach}
criticisms of Newton, is realized mathematically in terms of
invariance under active diffeomorphisms.  As has been
discussed in detail, the so called ``problem of time" in
classical and quantum cosmology is no more and no less than
the expression of this situation with respect to the notion
of evolution: the solution is that time evolution can only
meaningfully be talked about in terms of relationships between
physical degrees of freedom\cite{carlo-time,julian-gr,julian-qc}.

The only exceptions to this are the
asymptotically flat case, or other cases in which boundary
conditions are imposed, in which case the diffeomorphism
invariance is broken by the conditions imposed on the boundary.

I would like to put this in the following way: the mathematical
structure of general relativity in the cosmological case
forbids any possibility of an
interpretation in terms of operations or measurements made
by an imaginary observer ``outside of the universe."  If we
want to speak of an observer outside of the system we must
modify the mathematical structure of the theory in a way that
breaks the gauge invariance and so allows us to represent,
in a kind of idealized way, the possibility of measurements made
by an imaginary external observer.

We may then say that there should be a
problem with extending quantum
mechanics unmodified to the cosmological case, because the
theory does not satisfy the principle, just stated, that the
mathematical
structure of the theory should forbid the possibility of an
interpretation in terms of an observer outside of the system.
The mathematical
structure of quantum mechanics in fact does implicitly refer
to the existence of things outside the system.  Among these
are
the clock, measuring instruments and observer that give
meaning to the expectation values and their time evolution.
It then follows that if we could find an interpretation of the
theory in terms of measurements made by observers inside
the quantum system, the theory would still be unsatisfactory,
because the theory would allow us to formulate an
interpretation
in terms of an observer outside of the system.  This is true
by definition if we assume we can extend the formalism of
quantum mechanics to the cosmological case without modification,
because in that case the formal structures that are used in the
standard Bohr interpretation are still present.  Whatever
other interpretation was offered, these structures could then be
used as the basis of an interpretation of quantum cosmology
in terms of an imaginary observer outside of the universe.

This is absurd, and it is my opinion that it is not enough just to
refrain from doing this.  As in the case of general relativity,
we should require that {\it the
mathematical structure of quantum theory
must be modified in such a
way that there is no possibility of an interpretation of the formalism
in terms of an observer outside of the universe}.

I would like to raise this statement to the level of a
fundamental principle, which we may call the
{\it Principle of the absurdity of the possibility of an outside observer.}
I have discussed its
motivation and history in more detail elsewhere\cite{forabner,lotc,me-ste}.
Here I
would like to go on to sketch what modifications of quantum
theory might lead to a theory that satisfies it.

If we follow the example of general relativity we have a clue,
which is that the extension of the theory to the cosmological
case should correspond to the restoration of a gauge
symmetry, such that the standard theory, with fixed external
observers, is obtained by breaking it in some way.
What should this new gauge symmetry be?   Given what
has been said, it is clear that
it should arise from treating the observer and the
quantum system on an equal footing, so that the split between
them can be made arbitrarily.

Thus, our goal is not to eliminate the observer, it is
instead, to relativize
him.  We would like a formalism that allows us to divide the
universe arbitrarily into two parts, and call one part of it the
observer and the other the system.  We would like there to
be something like a gauge symmetry, that expresses the
arbitrariness of the split.  And, most importantly, to satisfy
the principle, we must do this in such a way that
it is impossible to construct a single state space that would
allow us the possibility of speaking in terms of  a description
of the whole system by an external
observer.

There is a natural way to do this, which is to associate
the operator algebras and their representations in terms
of Hilbert spaces not to the universe as a whole, but to
every possible splitting of the universe into
subsystems\footnote{To my knowledge, this proposal was
first made by Crane\cite{louis}}.
Thus, for every way of splitting the universe into two parts
we will have an associated state space and observable algebra.
Unlike quantum mechanics, there will not be one preferred
split, every possible split will be allowed.  Instead of conditions
that pick out which split is given an interpretation (analogous to
the preferred basis problem) we seek consistency conditions that
hold among the set of all state spaces and observable algebras.
These conditions will specify the ways in which the observations
made by each possible subsystem of the universe on the rest of it
may be consistent with each other.

Thus, our slogan is ``Not one state space and many worlds, but
one world, described consistently by many state spaces."

It is important for this point of view that we take the
statistical interpretation of the quantum state, according
to which the wavefunction is not something real, but is only
a representation of the information that the observer has
about the system\cite{statistical}.  We may note
that the mixing of quantum
and ordinary thermal statistics
in gravitational fields, and the fact that they are mixed
up by transforming between inertial and noninertial reference
systems, strongly suggests that quantum statistics are
really ordinary statistics which arise due to some non-local physics,
and are distinguishable from conventional thermal effects only by
special observers in special situations\cite{me-mixing}.  This point has
been discussed at length elsewhere, for the present, it is only
important to appreciate that there are reasons why the unification
of quantum theory and relativity might force us to accept the
statistical view of the
quantum state\footnote{Of course, there are claims of
the ``observability of the quantum
state".  However, these seem to be in fact statements of the
observability of non-classical or
non-local quantum effects, which are undeniably
real.}.

Further, we may note that there are clear arguments against
the possibility of an observer, in either classical or quantum theory,
making a complete measurement of the state of any system
that includes them\cite{nonself}.  These arguments reply on the
fact that a measurement requires a representation, within a
subsystem of the observer, of the information gained about
the system.  To make a measurement of something is then
to gain some
information, which is represented in the state of some degrees
of freedom inside the observer.  Thus, when an observer makes
an observation, its state must change, to record the information
gained. But if the observer is contained in the system, then
the state of the system changes consequently, as a result of
the measurement, from the measured state.  The observer may
attempt to again measure its state to ascertain the change, but
again the state must change, so that the
result is an infinite regress.
Moreover, for a complete self-measurement he would need an
infinite number of degrees of freedom, because there is a problem
of infinite regress, each time he observes himself he needs a new
place to put the information gained specifying his previous state.

Thus, as soon as we accept the point of view that the quantum state
corresponds in some exact way with the information that an
observer may gain about the system
as the result of an observation of it, we must accept also that no
such description can give a complete specification of the whole
universe, assuming that it also contains the observer.  The only
way to have a description which is both complete and
quantum mechanical is to use the fact that in fact the
universe is complex enough that there are many subsystems that
may be called observers, and to then postulate that the complete information
for describing the state of the universe is represented in a number
of Hilbert spaces, each of which represents the incomplete
information that
a particular observer may have about the
universe.

We might also note that taking the Hilbert space to correspond to
the interaction of an observer and a system brings the interpretation
of the formalism into complete correspondence with what we
actually do when we apply quantum theory to the real world.  For
in real practice, physicists describe different isolated systems with
different Hilbert spaces and algebras of observables.   One thing
that this interpretation accomplishes is thus to make the theory
correspond
directly to actual practice.  We no longer need to
imagine that all the actual applications of quantum theory are
to a greater or lessor extent approximations to some ``real
quantum description" in terms of some ideal ``Hilbert space of
the universe."

\subsection{The measurement problem and the consensus condition}

If we accept the basic idea of pluralistic quantum cosmology,
we no longer have to answer
questions about which split or which basis or which set
of histories corresponds to reality.  Instead, we have to answer
a new kind of question, which is what relationships may be
expected to hold between any two possible Hilbert space
descriptions, each giving an incomplete description of the world
arising from a different division of the universe into two
subsystems.  This becomes the key question for the theory;
we will see, indeed, that the whole of the theory,
and indeed, even its dynamics, may be expressed in terms of
these relationships.

The first questions that must be asked about these relationships is
whether the existence of different incomplete descriptions of the
same system can be consistent with the
principles of quantum mechanics, and in particular with the
uncertainty principle.  Further, does this point of view
shed any light on the measurement problem?
An argument of Rovelli suggests that the answer is yes\cite{carloqm}.
I would like now to review Rovelli's argument, as it suggests
the need for a certain consistency condition, which I
will call the {\it consensus condition.}

Suppose that there is a quantum system $A$ surrounded
by a two surface ${\cal S}_1$, which we may informally
call the ``box" that contains $A$.
Surrounding that box is second, ${\cal S}_2$,
which contains the
system $A$ and an observer, $O_1$. Outside of that is
a second observer, $O_2$.   The measurements the
two observers,
$O_1$ and $O_2$ make are recorded in the states in two Hilbert
spaces we may call  ${\cal H}_{{\cal S}_1}$ and
${\cal H}_{{\cal S}_2}$.

Now, let us assume that $O_1$  measures the
state of $A$, using a measuring instrument that
measures fields on the surface ${\cal S}_1$.  The
result is described differently by the two observers,
using the two Hilbert spaces.
The first observer's instruments register a definite value,
$\lambda_1$
which means that she will project the state that describes
$A$ into a definite state $|1> \in {\cal H}_{{\cal S}_1}$.
The second observer measures the combined state of
$A$ and the measuring instruments and observer
$O_1$, and learns only that as a result of their interaction
they are in some correlated state,
$\sum_i |i>_1 \otimes |i>_2 \in {\cal H}_{{\cal S}_1}$
where the states $|i>_i$ are the quantum states of the
first observer in which she has observed the system
$A$ to have eigenvalue $\lambda_i$.

The point is that there is no contradiction between these two
records of the observation.
The measurement of the second
observer ascertains that the combined system of $A$ and
the first observer is in a correlated state, corresponding to a
measurement having been made.  But he is unable to acertain what
the result of the measurement was.  In the meantime, the
first observer, having made the measurement, has observed
a definite value $\lambda_1$, and has reduced her description
of the system $A$ to the corresponding eigenstate.

This example shows that it may be possible to have different
Hilbert spaces which describe the results of observations made
by different observers, in a way that allows the different states
to be different from each other, but also consistent with each
other.  This example also shows us the way to
a condition of consistency
which must be imposed on the quantum states assigned to different
boundaries.  The state $|i_o>$ on the inner surface seen by
the observer must not be precluded by the state
$\sum_i c_i |i>_1 \otimes |i>_2$ seen by the second observer,
i.e. the ampitude $c_{i_0}$ cannot be zero.  We will call this
the {\it consensus condition}.  Stated formally, it means that
\f
Tr_1 \left [ \rho^1   [Tr_2 \rho^{1+2} ]  \right ] \neq 0
\ff
where $\rho^1$ is the statistical state the observer has that
represents his information about the system, $\rho^{1+2}$
is the statistical state describing the whole state of the
observer and system and $Tr_2$ means a trace over the
observers degrees of freedom, while $Tr_1$ means a trace
over the system's state space.

\subsection{Other consistency conditions}

There are two other consistency conditions that it is natural
to impose at this stage.  The first is that if the surface
is composed of two disconnected parts,  ${\cal S}_1$
and ${\cal S}_2$, we must have
\f
{\cal H}_{{\cal S}_1 \cup {\cal S}_2 } =
{\cal H}_{{\cal S}_1  } \otimes { {\cal S}_2 }
\ff
and, secondly, reversal of orientation is associated with
Hermitian conjugation, so that,
if $\bar{\cal S}$ is the same surface, with
orientation reversed,
\f
{\cal H}_{\bar{\cal S} } = {\cal H}_{{\cal S}  }^\dagger
\ff

The states associated
with ``the observer" then correspond to linear functionals of
the states that correspond to ``the system", which is natural
given that the state space is a description of their interface.

A simple objection may be raised to this, which concerns
the case in which the area of the surface is small compard
to the size of the whole universe. In this case the
surface may contain a small observer looking at a large universe,
but might it also not surround a small system being studied with
all the resources available in a large universe?
Thus, in the limit that the surface shrinks down, shouldn't one
of its associated Hilbert spaces get very big, while the other
becomes very small?  The answer to this is that were this to happen
it violates our {\it Principle of the impossibility
of an interpretation of a cosmological theory
by an observer outside of the universe.}   We must keep in mind
that the state space associated to a boundary
describes the information that an observer
on one side may know about the part of the world in the other
side.  As such, when the area of the surface decreases, the dimension
of the associated state space may decrease, representing either the
limited information capacity of the observer, in the case that the
``observer may be seen to be becoming small" or the decreasing
number of degrees of freedom of the ``system" in the case that it
may be becoming small.

However, while it may help to think this way, we should also
remember that there is no guarantee that the volume in one
region or the other is a good physical observable in quantum
gravity.

\section{Postulates of pluralistic quantum cosmology}

In the next section we will turn to the mathematical structure
which is available to implement these ideas.  For the present
we may summarize the results of the reasoning up till this
point in a list of postulates, that we would like a quantum
theory of cosmology to satisfy.

{\bf 1)  Definition of state spaces:}
A quantum cosmological theory associates to every
oriented three dimensional surface
$\Delta={\cal S} \times R$, with  $\cal S$ a compact two surface
without boundary,
a Hilbert space ${\cal H}_{\Delta}^{QG}$, which is a representation
of an algebra of operators ${\cal A}_{\Delta}^{QG}$.
These assignments of Hilbert spaces satisfy the conditions (4)
and (5).  Furthermore, for every cobordism $\cal M$, which
is a four manifold with boundaries
$\partial {\cal M}= \Delta \cup \Delta^\prime$ there is a linear
map,
${\cal M}_\Sigma : {\cal H}_{\Delta}^{QG}
\rightarrow {\cal H}_{\Delta^\prime}^{QG}$.

{\bf 2)   Consensus condition:}
A {\it pluralistic quantum state of the universe} is an
assignment of a statistical state $\rho_{{\cal S}\times R}$ into
each ${\cal H}^{QG}_{{\cal S}\times R}$, subject to the
{\it consensus condition}, which in diffeomorphism invariant
langauge is expressed as follows.  Let the surface $\cal S$
be of the form of two disconnected pieces
${\cal S}= {\cal S}_I \cup {\cal S}_{II}$.  Then we know
from (4) that
\f
{\cal H}^{QG}_{{\cal S}\times R}=
{\cal H}^{QG}_{{\cal S}_I \times R} \otimes
{\cal H}^{QG}_{{\cal S}_{II}\times R}.
\ff
The consensus condition is then the requirement that
\f
Tr_I \left ( \left [  Tr_{II} \rho_{I \cup II}   \right ] \rho_I
\right ) \neq 0
\ff

The interpretations of these objects are the following.  For every
$\Delta $ we may construct a four dimensional region of spacetime
${\cal W}$ such that $\Delta \subset \partial {\cal W}  $.
The region $\cal W$ may contain an
observer, together with clocks and measuring instruments that are
able to measure the values of fields induced on $\Delta$.  The algebra
of observables ${\cal A}^{QG}_\Delta$ then must have a classical
limit which corresponds to an algebra of fields in a classical
spacetime induced on a surface $\Delta$ (see {\bf 4}, below.)
A state $\rho_{{\cal S} \times R}$ then corresponds to the information
that an observer in $\cal W$ may have about the fields induced on
$\Delta = {\cal S} \times R$.  These, together with knowledge of
the constraints and dynamics,  may then be used to infer
information about the values of fields in the region on ``the other side of
$\Delta$".

For example, if ${\cal S}= S^2 \times R$ whose interior
is ${\cal W}= S^3 \times R$, then we may say that an observer who lives
in the ``world-tube" ${\cal W}$ is able, by means of measurements of fields
on the boundary $S^2 \times R$, to gain information about the
universe external to the boundary.  This information is then recorded
in the value of a statistical state $\rho_{S^2 \times R}$.

{\bf 3) No interpretation in terms of external observers:}
${\cal H}_{{\cal S} \times R}^{QG} \rightarrow C$, the complex
numbers,  in the limit that
the two surface $\cal S$ shrinks to a point.

{\bf 4)  Incorproration of quantum gravity:}
${\cal A}_{\Delta}$ must
have a subalgebra ${\cal A}_{\cal S}$, which includes observables
that measure the spacetime metric $g_{ab}$ and left handed
spin connection $A_A^{AB}$ pulled back into the
two surface ${\cal S}$.  (These will be denoted $h_{\alpha \beta}$
and $a_{\alpha}^{AB}$ respectively.)

{\bf 5) The Bekenstein condition:}
Let ${\cal H}_{{\cal S}, h}$ be the subspace of
${\cal H}_{{\cal S} \times R}^{QG}$ containing eigenstates
of the two metric on $\cal S$ with eigenvalues denoted by
$h$.
Then, if $A(h)$ is the area of the two surface metric we
require,
\f
dim ({\cal H}_{{\cal S}, h}) = e^{c A(h)/l_{Pl}^2}
\ff
where $c$ is a fixed dimensionless constant.

 Finally, we require an axiom of correspondence with linearized
quantum field theory, which may be formulated as follows,

{\bf 6)  Correspondence with linearized QFT:}
Let us choose a cosmological constant $\Lambda$ (which is
assumed one of the paramters of the theory), an area $A_0$
and an ultraviolet cutoff $r$ such that
\f
l_{Pl}^2 << r^2 << A_0 << \Lambda^{-1/2}
\ff
Then, let ${\cal H}^{lin}_{\Lambda , A, r, \epsilon }$ be the
Hilbert space of
linearized gravitons (and other massless fields, if desired) restricted
to a region of
DeSitter spacetime with cosmological constant $\Lambda$ bounded
by a three boundary $S^2 \times R$, with the metric induced on
the two sphere being a two sphere metric with area $A$ by the
condition that the linearized fields vanish at the boundary.  Further,
in the reference frame in which the boundary is static two conditions
are put on the spectrum.  First, the energy
of the individual gravitons and other massless quanta
must be less than the inverse cutoff $r^{-1}$.  Second, the total
energy (defined in terms of the quasi-local surface integral
$H=\int_{S^2} \kappa $) must be bounded by
\f
E < \epsilon \sqrt{A}
\ff
with epsilon much less than one.  Let
${\cal A}^{lin}_{\Lambda , A, r, \epsilon }$ be an algebra of
observables associated with this space.

The meaning of this space and algebra is they define the linearized
limit of any quantum theory of gravity with small cosmological
constant, restricted to the interior of a static surface.
Two cutoffs are needed, one to limit the energy of any
single graviton to less than the Planck scale, the other to limit
the total energy to be much less than that at which gravitational
binding energy, and the collapse to a black hole, would be relevant.

For the quantum theory of cosmology to have an acceptable
linearized quantum field theory limit we require that here
exists a family of maps,
\f
{\cal N}_{\Lambda , A, r, \epsilon } :
{\cal H}^{lin}_{\Lambda , A, r, \epsilon } \rightarrow
{\cal H}_{S^2 \times R}^{QG}
\ff
\f
{\cal N}_{\Lambda , A, r, \epsilon } :
{\cal A}^{lin}_{\Lambda , A, r, \epsilon } \rightarrow
{\cal A}^{QG}_{S^2 \times R}
\ff
such that
\f
<{\cal N} \circ \Psi |{\cal N} \circ
{\cal O} |{\cal N} \circ  \Psi >^{QG} =
<\Psi |{\cal O} |\Psi >^{lin} + O(r/l_{Pl} ) + O(\epsilon ) +
O(l_{Pl}^2 /A )
\ff

This guarantees that there is a subset of the physical states and
observables whose physical expectation values reproduce the
predictions of the linearized theory.

To proceed to investigate the reasonableness of these postulates
we must ask if it is consistent with the physical interpretation of
both general relativity and quantum theory, especially given the
special problems associated with the notion of time.  This is the
subject of the next section.

\section{The problem of time}

One advantage of the proposal just made is that it may allow a new
possibility for the resolution of the problem of time, which has
plauged attempts so far to construct a quantum theory of
cosmology.

To introduce the basic idea, let me recall some of the proposals to
resolve the problem of time in quantum cosmology.  One, due to
DeWitt\cite{bryce-time} and developed and
advocated by Rovelli\cite{carlo-time},
is to use the
Heisenberg picture and construct physical operators that
describe correlations between certain degrees of freedom,
associated to clocks and other degrees of freedom.  In the
classical theory, it is certain that there are a large set of physical
observables (that is functions on the kinematical phase space
that commute with the constraints under the Poisson brackets)
which are what Rovelli calls ``evolving constants of motion."
It then may be that operators can be constructed in the quantum
theory that have the same meaning (although, of course there are
daunting technical problems in doing this for a realistic field theory
rather than for an integrable system.)

However, even if these technical problems could be overcome,
there is an important conceptual problem, which is how the
expectation values of these operators are to be correlated with
observers made by us observers inside the universe.
If we could imagine that there are observers outside of the
universe, who make observations of the universe, and are able
to many times prepare and measure the state of the whole
universe, they would
be able to verify whether or not the expectation values of these
``evolving constant of motion operators" in a particular physical
state are in accord with their observations.  But, this is something
we, as observers in the universe, are unable to do.

If we insist on our principle that there should be no possibility
of speaking in terms of an observer outside of the universe, then
Rovelli's proposal, even if it works technically, would not lead
to predictions that could be checked by us observers inside
the universe.

The same situation holds in other proposals concerning the
problem of time in quantum cosmology such as the proposal
of Page and Wootters\cite{pagewooters},
and the proposal of \cite{julian-qc}.  In the
first, one imagines that the Hilbert space is a direct
product
\f
{\cal H}= {\cal H}^{clock} \otimes {\cal H}^{other}
\ff
of a space representing a clock degree of freedom and
a space representing other degrees of freedom.
Physical quantum states states describe correlations
between certain degrees of freedom considered to be
clocks and the other degrees of freedom, coded in the
entanglement of the state, as in
\f
|\Psi > = \sum_i c_i |t_i>^{clock} \otimes |\chi_i >^{other}  .
\ff
These correlations are understood to be induced by
the constraints, which impose the dynamics.  The different
possible physical states then correspond to the different
sets of correlations that are possible between the clock
and the other degree of freedom, given the dynamics.

Again, we have the problem that the predictions of the existence
of such correlations could only be detected by an observer
outside the system.  Even if we accept the many worlds
point of view, which Page and Wooters seem to do, an
observer inside the system can  only verify the presence
of one branch, but they cannot give meaning to the coefficients
$c_i$ nor they can tell if the theory gives the right relative weights
to the different correlations.

Conversely, in the Barbour picture, the notion of a
single clock degree
of freedom is given up in favor of the more realistic picture that
time is measured from the coincidences among all the degrees of
freedom in a complex universe.  Instead,
the ``wavefunction of the universe" is interpreted to give directly
the relative frequency for the occurance of instantaneous
configurations in a great collection of such configurations.  The
collection of such ``moments" is taken to constitute reality; any
impression of time comes from the existence of memories, records
and what Barbour calls ``time capsules" in complex
instantaneous configurations.

The problem is again that only an observer outside of the universe
can give a meaning to these relative frequencies for ``moments"
in Barbour's conjectured reality.

Of course, all of these proposals allow that classical general
relativity coupled to quantum fields is
recovered if the state is appropriately ``semiclassical".  But
while a necessary condition for an interpretation, this is not
sufficient, as a good quantum theory of cosmology should in
principle give predictions for all states, not only those that
are semiclassical.  Otherwise, its predictions may not differ
from those of semiclassical gravity, perhaps together with
the addition of some particular boundary condition.

 From the present point of view, the problem which all these
proposals have in common can be resolved
in the following way.  The physical quantum state is now
associated to a three dimensional surface, which is understood
to represent the evolution in time of the two dimensional
boundary of a part of the universe.   The description can
be completely invariant, with respect to diffeomorphisms both
in and inside of the surface.  Thus, time must be understood in
terms of correlations between degrees of freedom inside the
surface, whether these are expressed in Barbour's, Page and
Wooter's or Rovelli's pictures.  But there are in fact observers
outside of the system, they live just outside the boundary.
Furthermore, if we use the statistical interpretation of the
quantum state, then the statistics predicted from the
expectation values of operators are interpreted in terms of
the relative frequency probability  for what correlations between
clocks and observers they will see if they observe the system
on their surface.

For example, if the state is of the form (15) then the $|c_i|^2$
are the probability that if they observe the system they will
find that the clock reads a time $t_i$ while the rest of the
system is in a state $|\chi_i >$ (assuming for simplicity the
trivial inner product.)  They can in principle check the
predicted probabilities by
preparing the part of the universe inside the boundary many
times in the state $|\Psi>$ and repeating the experiment.

Suppose our observers
do the experiment in sequence many times, will
they record a sequence of times that are in cronological order?
Will the times they measure for the clock in the system be
synchronized with clocks they may carry outside of their
boundary?  To answer such questions we cannot just rely
on the quantum description of the observers themselves.
Such questions can only be answered in the quantum theory
by enclosing the observers and the system inside a second
boundary ${\cal S}^\prime \times R$ and constructing a Hilbert
space to represent the combined system of the first system and
its observer.  The state of this combined system then describes
the information that a second observer, outside the second
boundary can have about it.  This observer can make observers
of the combined system of the first observer plus the first system,
and by doing so answer questions like these.

In this context as well, the consensus condition (3)  is essential
to guarantee that the probabilities actually seen by the first
observer agree with the $|c_i|^2$ in the expansion of the state
seen by the second observer.
For, by the theorem of
Finkelstein\cite{finkelstein-prob} and Hartle\cite{hartle-prob}
it guarantees
that in the limit of a large number of observations, these two
probabilities will agree.  Thus, in the present context, this
theorem plays an essential role, but one rather different than
in the many worlds interpretation.  Rather than giving a meaning
to probability it guarantees that the probabilities for events
seen by the different observers are consistent with each other.

Thus, it may be said that this interpretation is giving meaning to
Bohr's statement that the split between the classical and quantum
world can be made arbitrarily\cite{bohr}.  For it allows us to consider
a description that is given by specifying simultaneously the states
associated with all such possible splittings.  It is very interesting
that to use this interpretation to resolve problems in quantum
cosmology such as the problem of time we must use some of the
results associated with the many worlds interpretation.  However,
as in the case of the theorem of Finkelstein and Hartle, their role
is transformed from giving a meaning to an interpretation in
terms of one state and many worlds, to ensuring the consistency of
an interpretation in which many states describe many partial
observations of one world.

\section{The classical limit and the timelike initial data problem}

If the picture we are describing is to work, it must have a classical
limit which makes sense in terms of classical general relativity.
We must then ask whether there is a formulation of classical
general relativity in the spacially compact case in
which the theory may be described in terms
of fields induced on a timelike surface $\Delta={\cal S} \times R$.
This is equvalent to posing the
{\it timelike initial data problem} \cite{timedata}.
This is the question of whether there is data, subject to constraints,
that may be given on such a surface $\Delta$, such that the three
metric given on $\Delta$ has signature $(-,+,+)$ and such that
the data determines a solution to Einstein's equations, at least
in a neighborhood of the surface.

The timelike initial data problem, along with its cousin, the
null initial data problem, is not as well studied as the conventional
initial data problem in terms of spacelike surfaces.  But, it has
often been observed that the Hamiltonian formalism does not
require that the surface on which the fields and momenta are
derfined be spacelike.  Thus, at least
formally, one can construct a formalism for first order evolution
of data, subject to constraints, from a timelike initial data surface.
In this formalism, momenta correspond to derivatives off of the
surface, and there are constraints, which are of the same form
as the usual constraints, with certain changes of sign due to the
change in the signature of the induced three metric.

This formalism will be discussed in detail
elsewhere\cite{inprep}.  For the
present, we need only know that at the formal level the
Hamiltonian formalism can be translated to a timelike
surface.  It is also interesting to note that in this case
it is the diffeomorphism constraints
that induces the correlations between degrees of freedom on the
surface that might be used to measure time and other degrees of
freedom. The role of the Hamiltonian constraint is then to develop
the solution into the interior, and thus establish correlations between
fields on the surface and fields in the interior.

\section{Incorporating topological quantum field theory
and quantum gravity}

Now that we have seen that the problem of time, as well
as the classical limit, seem to pose no problems of principle,
we may turn to the question of trying to realize the
postulates made in section 5 in terms of a concrete
mathematical construction.

The proposal made above relies on an apparant coincidence,
which is that
both the Bekenstein
bound and the point of view about quantum cosmology proposed
in the section 4 lead to the same conclusion, which is that
quantum gravity, in the cosmological case, may
be defined in terms of state spaces
attached to three dimensional ``timelike"
surfaces which represent the
information that observers who measure fields induced on the
surface may have about the degrees of freedom
in its interior.  Thus,
it is very natural to synthesize the two lines of development,
which is what we have done in formulating these postulates.

The question that must now be posed, however, is how
these postulates are to be realized in a concrete theory.
In the following sections I will present one proposal for
how to do this, which is based on a further
apparent coincidence.

In order to proceed the main question that must be answered
is how the finite dimensional state spaces, which are the
eigenspaces of the two metric operators are to be realized.
The hope to realize this and the other conditions comes from
an apparent coincidence of two results, one from non-perturbative
quantum gravity and one from topological quantum field theory.
The first is the discovery that in non-perturbative quantum gravity
the operators that measure areas of regions, $\cal R$ of
any two dimensional surface $\cal S$, which may be
denoted $A[{\cal R}]$, have discrete spectra.
To say this more precisely, the basic result is that
the quantum states of a diffeomorphism invariant
quantum field theory in a region $\Sigma$ are given by the
diffeomorphism classes of
the spin networks in $\Sigma$ \cite{volume,spinnet-us}.
The spin networks
are the eigenstates of $A[{\cal R}]$ and the corresponding
eigenvalues are given
by the spins of the edges of the network that intersect
the surface at points in the region $\cal R$.  Thus,
if $|\Gamma >$ is the quantum state associated to the
spin network $\Gamma$ in $\Sigma$ and $y_\alpha^{\Gamma, {\cal R}}$
are the points at which the spin network meets the region
$\cal R$ of the boundary, and $j_\alpha$ are the spins of
the edges that meet
boundary, we have
\f
\hat{\cal A}[{\cal R}] |\Gamma > = l_{Pl}^2 \sum_\alpha
\sqrt{j_\alpha (j_\alpha +1 )} |\Gamma >
\ff

One way to measure the two metric, up to diffeomorphisms, is
to measure the $A[{\cal R}]$ for all the
${\cal R} \subset {\cal S}$.  We see then that the two
metric can be considered to have a discrete spectrum.
The simultaneous eigenspaces of the $A[{\cal R}]$
are labled by punctures $y_\alpha$ marked by spins
$j_\alpha$ which label representations of $SU(2)$.
Thus, the eigenspaces ${\cal H}_{{\cal S}, h}^{QG}$, which
should have finite dimension, should more precisely be
labled ${\cal H}_{{\cal S}, y_\alpha , j_\alpha}^{QG}$.

As a result, there are no problems with continuum measures in
the statement of postulate {\bf 5}, so that we may write,
\f
{\cal H}_{{\cal S} \times R}^{QG} = \oplus_n \oplus_{j_\alpha}
{\cal H}_{{\cal S}, y_\alpha , j_\alpha}^{QG}
\ff

This is a step, but it does not yet explain to us how to
construct state spaces that are eigenspaces of these operators
that have finite dimension that grows exponentially with the
area.  To see how to do this, we must turn to some basic
results of topological quantum field theory.

Before explaining how this problem is solved, though,
we must divert to discuss
a remarkable observation of Louis
Crane, which is that the axioms of topological quantum field
theory in three dimensions are exactly what is needed to
guarantee the consistency of a many-Hilbert space
reformulation of quantum theory of the type we have
been discussing\cite{louis}.     This is because by the axioms of
Atiyah\cite{atiyah}, TQFT's automatically assign state spaces to boundaries,
so that the conditions (4) and (5) are satisfied.  Moreover, the
posulate {\bf 3)} is {\it automatically} realized as a result
of the axioms.
For these and other reasons, described in his papers, Crane
has advocated the construction of a quantum theory of
gravity as an extension of topological
quantum field theory.
The proposal I will shortly make is indeed an attempt to
realize this program\footnote{Other attempts to realize
quantum gravity as an extension of TQFT are described
in\cite{louis,louisdavid,louisigor,baezdolan,rjl,jb}.}.

Furthermore, topological quantum field theory naturally
extends to the case of two dimensional
surfaces with points marked by
representations of groups\cite{louis2d3d} and spin networks play
a natural role in this formulation.  Among the axioms of
$TQFT$ one finds that a three dimensional $TQFT$
associated to a quantum group $G$ is a functor that
assigns\cite{atiyah,louis}:

\begin{itemize}
\item  to every two dimensional surface $\cal S$ with
$n$ punctures $y_\alpha$, $\alpha=1,...,n$ labled with
representations $j_\alpha$ of $G$ a finite dimensional
Hilbert space ${\cal H}^G_{{\cal S},j_\alpha}$.

\item  to every three manifold $\Sigma$ with imbeded
$G$-spin network $\Gamma$, such that
$\partial \Sigma = {\cal S}$ and $\Gamma$ intersects
$\cal S$ in the $n$ labled points $y_\alpha$, with the
line intersecting $y_\alpha$ being labled by the same
representation $j_\alpha$, a state
$|\Gamma , \Sigma > \in {\cal H}^G_{{\cal S},j_\alpha}$.

\end{itemize}

Note that this is defined to be a functor from the category
of manifolds, with cobordims as maps to the category of
Hilbert spaces, with linear maps as the maps.  This
guarantees that the conditions (4) and (5), as
well as postulate {\bf 3)}  are automatically
satisfied.

Thus, it seems that topological quantum field theories
are exactly the objects we need to construct the
finite dimensional state spaces that represent
eigenspaces of ${\cal H}_{{\cal S}, h}$.  More precisely,
if we use the coincidence that spin networks and points
marked with representations of a group appear in both
$TQFT$ and quantum gravity
we may then make the hypothesis that there is a
topological quantum field theory associated with $SU(2)$,
to which we may identify these spaces, so that
\f
{\cal H}_{{\cal S}, y_\alpha , j_\alpha}^{QG}
= {\cal H}_{{\cal S}, y_\alpha , j_\alpha}^{TQFT}
\ff
There is a natural candidate for this TQFT,
which is the Chern-Simons theory associated with the
group $SU(2)$.    The variable there, which is an $SU(2)$
connection, is exactly what we need to represent
$a_\alpha^{AB}$, the pull back of the Ashtekar connection
to $\cal S$.
If we are to  try to realize this  hypothesis, however,
there are three
questions we must answer.
\begin{itemize}

\item Can we give an interpretation, in terms of
the parameters of quantum gravity, to the coupling constant,
or level, $k$ of the Chern-Simons theory?  This involves two
related puzzles, first that it
seems to be ordinary $SU(2)$ spin networks that play
a role in non-perturbative quantum gravity, while it is the
q-deformed networks that are relevant for TQFT, and second that
the commutation relations of the two theories are different.

\item  Can the constraints of quantum gravity be represented, or
its solutions expressed, in terms of the states of the
Chern-Simons theory?

\item  Is the Bekenstein bound satisfied?

\end{itemize}

It turns out, as I will describe in the next section, that there
are natural answers to the first two questions and that,
furthermore, the answer to the third question is yes.

\section{A proposal}

 From the side of quantum gravity, we seek a theory that provides
a representation for the observables that will be induced in
a three dimensional timelike surface, $\Delta = {\cal S} \times R$,
where $\cal S$ may be considered an intersection of $\Delta$
with a surface of constant ``time".  We may then, to make a
connection
with the standard $3+1$ canonical formalism consider the algebra
of observables induced in the two surface $\cal S$.
These include the pull backs into $\cal S$ of the
curvature two form $F_{ab}^{AB}$ and the dual of the
densitized from field $E_{ab}^{AB} = \epsilon_{abc} \tilde{E}^{c AB}$
 which I will denote by $f^{AB}$ and $e^{AB}$,
respectively\footnote{$A,B,...$ stand for
two component spinor indices.  We will also choose conventions
in which $f^{AB}$ and $e^{AB}$ have dimensions of $length^{-2}$
and $length^o$, respectively.}.  We may
note that these all commute with each other under the standard
Poisson
brackets of general relativity.

In ref. \cite{linking} I studied the case in which $\Delta$ was a
boundary
of spacetime, to which we needed to associate particular boundary
conditions
as well as a boundary term in the action.  There I studied a
particular
condition called the ``self-dual" boundary condition, in which
\f
f^{AB}= {2 \pi \over k G} e^{AB}
\ff
where $k$, was required by gauge invariance to be an integer, and
was
as well related to the parameters of general relativity by
\f
k = {6 \pi \over G^2 \Lambda} + \alpha
\ff
where $G$ and $\Lambda$ are Newton's constant and the
cosmological constant,
respectively and $\alpha$ is a $CP$ violating parameter coming from
an
$\int F^{AB} \wedge F_{AB}$ term in the action.

This has the effect of inducing a topological quantum field theory
in the boundary, which was in fact $SU(2)_q$ Chern-Simon theory,
with
$k$ the level, or quantum deformation parameter.  This is possible
because an additional effect of imposing the boundary condition (19)
is that the algebra of observables in $\cal S$ was deformed, so that
the connection one form $a_\alpha^{AB}$, which is the pull back of
the Ashtekar connection $A_a^{AB}$ into $\cal S$ satisfies,
\f
\{ a^{AB}_\alpha (\sigma ) , a^{CD}_\beta (\sigma^\prime ) \} =
{2 \pi \over k} \epsilon_{\alpha \beta } \delta^2
(\sigma , \sigma^\prime )
(\epsilon^{AC}\epsilon^{BD} + \epsilon^{AD}\epsilon^{BC})
\ff
which are in fact the Poisson brackets of Chern-Simons theory.

As a result, a Hilbert space that represents this
algebra of observables, subject to the condition (19)
is given, for each choice of the boundary $\cal S$ by
the direct sum,
\f
{\cal H}^{gr}_{\cal S} = \oplus_{n=1}^\infty \oplus_{j_1 , ... , j_n}
\int d^2y_1 ... d^2 y_n
{\cal H}^{CS, k}_{{\cal S}, y_\alpha , j_\alpha}
\ff
where $y_\alpha$  with $\alpha = 1, ...n$ are the positions of
$n$ punctures on the surface $\cal S$,
the $j_\alpha$,are $n$ labels of representations
of $SL(2)_q$ (or quantum spins) located at $n$ punctures on the
surface $\cal S$.  By standard constructions in TQFT there is a finite
dimensional Hilbert space associated with each choice of
${\cal S},n ,  j_\alpha $ and $k$.

A question that was left open in \cite{linking} was whether the
diffeomorphisms of the two surface $\cal S$ can be imposed as a
gauge
symmetry.  There seem to be technical obstacles to doing so, but
there is also reason to believe they may be overcome.  In this case
the state space associated to the surface would simply be
\f
{\cal H}^{gr}_{\cal S} = \oplus_{n=1}^\infty \oplus_{j_1 , ... , j_n}
{\cal H}^{CS, k}_{{\cal S} , j_\alpha}
\ff
because, of course, the Hilbert space of the TQFT does not depend on
the positions of the punctures.

In reference \cite{linking} the conjecture was then made that the
algebra of observables on the boundary was sufficient to label the
quantum states of the interior.  While this is not yet shown, there
is some evidence for it, which was discussed there.  If this is the
case, then the physical state space of quantum gravity, in the
presence of the self-dual boundary conditions, would then have
the form (22).

In the case we are discussing here, the two
surface plays a somewhat different role.  Rather then being a
boundary
of space, at which certain boundary conditions may be imposed, it is
now to be thought of as simply an arbitrary two dimensional surface
in
space.  The question is whether any of the structure discovered for
this particular set of boundary conditions is at all relevant to this
case.

To investigate this question, let us ask what replaces the self-dual
condition (19) in the general case?  We may recall that, in the case that
the field equations (or at least their pullback,
into $\cal S$) are satisfied, there is a relationship between the fields
$f^{AB}$ and $e^{AB}$ in $\cal S$, which is given by
\f
f^{AB} = \left (\Psi^{ABCD} +  {G \Lambda \over 3}
\epsilon^{AB} \epsilon^{CD} \right ) e_{CD}
\ff
where $\Psi^{ABCD}$ is a totally symmetric spinor, representing the
spin-two degrees of freedom of the gravitational field.  This
relationship can be seen both from the $CDJ$ formalism\cite{CDJ} and the
Newman-Penrose equations\cite{tedroger}.  We may note that this equation
constitutes
in fact the general solution to the Hamiltonian and
diffeomorphism constraints, or equivalently the frame field
field equations, pulled back into any three surface that contains
$\cal S$.

This equation can also be understood to be an expression of the
hamiltonian and diffeomorphism constraints.  As pointed out by
Reisenberger\cite{mikerconstraints},
one can express the constraints coming from
the CDJ action by the statement that there must exist a symmetric
spinor $\Psi^{ABCD}$ such that (23) is satisfied.  This may have
some advantages over the standard, Ashtekar, form of the constraints,
as its derivation makes no assumption of the invertibility of the
metric.

How are we to represent this relationship quantum mechanically? It
is
clear that what the equation is saying is that in general the
two fields $e^{AB}$ and $f^{AB}$ are free to some extent to vary
independently, as long as they are related by (24). This means that
they can differ by the action of two terms, there is a contribution
to $f^{AB}$ which is proportional to $\Lambda$ times $e^{AB}$ and
there is a second term which is proportional to the action of a spin
two field.

To translate this quantum mechanically let us note that in the case
that the spin two field is absent they are strictly proprtional to
each other.  In this case, how is the equation
(19) represented quantum mechanically?
In the loop representation, $e^{AB}$ will
be represented by a set of punctures, which are the intersections of
a spin
network $\Gamma$ with the two surface $\cal S$.  These punctures
are
labled by representations of $SU(2)_q$, which are those carried by
the
lines of the spin network at the intersections.  If
$f^{AB}$ is, in this
limiting case, to be proportional to $e^{AB}$, then it must also be
represented by the same set of punctures, labeled by the same
representations.
This is what happens when we impose the self-dual condition (19),
as described in \cite{linking}.  The
result
is that, under these conditions, when we measure the metric of the
two
surface (and hence $e^{AB}$), we pick out an eigenstate of the
operators
that measure elements of area in the surface, and this picks out a set
of
punctures and representations.  Because of the proportionality of
$e^{AB}$ and $f^{AB}$, the result is that this picks out a two
dimensional quantum field theory, which is $SU(2)_q$ Chern-Simons
theory with that choice of punctures and representations.  The states
of this theory represent the freedom that the connection,
$a^{AB}_\alpha$,
still has, once the condition (19) is imposed.

Now let us consider the case that the spin two field $\Psi^{ABCD}$ is
non-zero, so that the self-dual condtion (19) is relaxed to (24).  In
this case we may represent the situation as follows.  The
representation
of the metric observables $e^{AB}$ should be the same, so these
will have eigenstates, associated with the intersections of spin
networks
with $\cal S$, which are hence labled by choices of
punctures, $y_\alpha$ and representations $j_\alpha$.
How are we to represent the degrees of freedom of the
connection $a^{AB}_\alpha$, now subject to the constrant (24)?  Let
us represent this as before by the Hilbert space of $SU(2)_q$
Chern-Simons theory, with the same set of punctures as before.
This is in agreement with (24), which tells us that
in regions over which the integral of $e^{AB}$ vanishes,
the integral of $f^{AB}$ must vanish as well.  However, when
it is allowed to be nonvanishing, let us
allow for the connection to vary independently by letting the spins,
which we will call $l_\alpha$ at these punctures vary.

Thus, the
degrees of freedom of the quantum theory will be punctures,
$y_\alpha$, each of which is labled by two spins, $j_\alpha$
and $l_\alpha$ corresponding to the separate metric and connection
degrees of freedom.  However, the metric and connection are not
completely free to vary independently, because they are constrained
classically by (24).  Since $\Psi^{ABCD}$ is a spin two field, this means
that quantum mechanically, each $l_\alpha$ must either be
proportional
to the corresponding $j_\alpha$, or must be in the decomposition of
the
multiplcation of $j_\alpha$ with the spin-two representation.
That is, the quantum mechanical version of (24) may be hypothesized
to be,
\f
l_\alpha \in ( 2 \otimes j_\alpha , j_\alpha )
\ff
(Of course $j_\alpha$ is already contained in the decomposition
of the product of itself with spin $2$, but for clarity, because it
comes from a separate term, we list it separately.)

The result is that the state space of quantum gravity, associated with
the two surface $\cal S$ now becomes,
\f
{\cal H}^{QG}_{\cal S} = \oplus_n \oplus_{j_\alpha }
{\cal H}^{QG}_{{\cal S}, j_\alpha}
\ff
where the ${\cal H}^{QG}_{{\cal S}, j_\alpha}$ are the eigenspaces of
${\cal A}[{\cal S}]$, the area of $\cal S$, with eigenvalues
$\sum_\alpha \sqrt{j_\alpha (j_\alpha +1 )}$ (neglecting corrections
in $1/k$ \cite{sethlee-qnet}.)  Each of these eigenspaces
is then further decomposed,
\f
{\cal H}^{QG}_{{\cal S}, j_\alpha}=
\oplus_{l_\alpha \in ({2} \otimes {j}_\alpha ,j_\alpha )}
{\cal H}^{CS, k}_{{\cal S} , l_\alpha}
\ff

Having specified the state space of the theory we must now discuss
the role of the parameter $k$.  We may first note that $k$ must
be considered to be a parameter of the theory as it comes into the
specification of the observables algebra, in (21).  We can see from
that equation that $k^{-1}$ not vanishing indicates that there is
a kind of anomolie in the theory, so that the classical commutation
relations have been modified so that all components of $A^{AB}_a$
no longer commute.  As is shown in \cite{linking} this turns out
to be related to the necessity to frame the loops that go into
the spin network constructions, so that a diffeomorphism
in $\cal S$ that has the effect of twisting a line of a spin
network (or rotating a puncture) has the effect of multiplying
the states of the theory in the spin network basis by a phase.

We may note that there are reasons to believe that this is a real
effect in the case of non-vanishing cosmological constant,
which comes from the attempts to define states in the
loop representation that would correspond to the loop transform of
the
Kodama state\cite{kodama}.  Since the anomolie is  proportional to
$k^{-1}$, it is consistent to take $k \approx \Lambda^{-1}$, as
equation (20) asserts.  To fix the value of $k$, we can use the
arguments from \cite{linking}, which apply, however, to the case that
$\Psi^{ABCD}$ is constrained to vanish, yielding eq. (20).
Alternatively, we can use arguments coming from the attempts
to define the path integral involved in taking the loop transform
of the Kodama state, which lead to the same
conclusion\cite{ls-cs,BGP-cs}.
It would,
however,
be better to have a more general argument, independent of either a
restriction of the degrees of freedom at the surface, or the choice
of a particular state.  Thus, at present, it is perhaps best to
take (20) over to
the present case as a working hypothesis.

Further, we may note that as (24) is the general solution to the
Hamiltonian
and diffeomorphism constraints, what we have expressed quantum
mechanically
through (25) is in fact a parametrization of the space of solutions
of the constraints.  Further, the fact that the frame field has been
expressed in terms of spin networks (leading to the punctures)
means
that the Gauss's law constraint has been solved.  We may then take
(27) to be a description of the physical state space of the theory.

Finally, we may note that the Bekenstein bound is in fact satisfied
in this case, at least in the limit of large $k$, or small
$G^2 \Lambda$.  The bound will hold
(in the limit $k \rightarrow \infty$)
if there is a fixed constant
$c$ such that
\f
\ln dim {\cal H}^{QG}_{{\cal S},j_\alpha} \leq c
\sum_\alpha \sqrt{j_\alpha (j_\alpha +1 )}
\ff
for all choices of punctures and labels.  However, in the same limit
the dimension of ${\cal H}^{QG}_{{\cal S},j_\alpha}$ is just the
product of all dimensions of
the allowed representations of $SU(2)$ at each
puncture\cite{witten-tqft}.  Thus, we have,
\f
\sum_\alpha \sum_{l\in (2 \times j_\alpha , j_\alpha )}(2l+1 )
\leq c
\sum_\alpha \sqrt{j_\alpha (j_\alpha +1 )}
\ff
It is not difficult to discover that this will be true for all
$c \geq 28/\sqrt{3}$.

We have thus achieved our goal, which is to postulate a starting
point
for a quantum theory of gravity that is consistent with
the results we want to keep from previous investigations, including
the kinematical structure of the loop representation.  Further,
we have the theory in a language which is suitable for pluralistic
quantum theory and, using the connection between spin networks,
as they represent eigenstates of the area operator in quantum
gravity,
and spin networks as they label states in topological quantum field
theory,
we have found a way to naturally implement the Bekenstein bound.
This
means that we have a field theory, associated to two dimensional
surfaces,
that implements the holographic hypothesis, and implements
naturally
the condition that the subspaces of the state space that are
eigenspaces of the metric of the boundary are finite dimensional,
with the dimension of the boundary growing exponentially with the
physical area.

\section{Conclusions}

It need hardly be said that the present theory is put forward
with all due reserve.  What I have described here is only
a sketch of a research
program.  There are several directions in which the proposals outlined
here might be pursued.  The theory sketched in the previous section
needs to be developed by the introduction of additional observables,
correspoding to components of the metric, connections and curvatures
which are not tangent to the surfaces.  It is possible, following
the results of section VIi of \cite{linking},
that this will enable the full tangle
algebra of Baez \cite{john-tangle} to be represented.

At the same time, one may try to sharpen the system of postulates
proposed here to a system of axioms that could be taken as the
foundations of a quantum theory of gravity.  The question in this
case is whether a system of axioms can be found which allows
both classical general relativity and quantum field theory on
fixed backgrounds to be derived in the appropriate limits.

In this respect, a
very interesting possibility is that Jacobson's recent
construction\cite{ted-new}
can be used to derive the Einstein equations directly
from thermodynamics
together with a set of postulates about the fundamental
quantum theory, such as those proposed here.
According to Jacobson's construction all that
is needed to derive the Einstein's equations as the classical
limit of a quantum theory of gravity are three assumptions:
1) the laws of thermodynamics 2)  the recovery of linearized
quantum field theory in restricted regions whose curvatures
are small in Planck units and 3)  the connection between
entropy and surface area.  The idea is that  2) may be guaranteed
by the last postulate while  3) is guaranteed by the Bekenstein
condition (8).  We may expect that
the laws of thermodynamics are to be recovered
from the usual statistical assumptions, given that the use of
ordinary, relative frequency statistics is guaranteed by the
consensus condition.  Whether this
can be carried out in detail is a very interesting question; if it can
then the conclusion would be that in a quantum theory of gravity
the fundamental axiom that introduces gravitation is the connection
between information and entropy, from which Newton's laws of
gravity would follow as consequences of the recovery of general
relativity in the classical limit.  This reverses the conventional
point of view in which the law of gravitation, expressed in terms
of Einstein's equations is taken to be fundamental, with the
connection between information and surface area a consequence.

In this respect, the key result from canonical quantum gravity
that is being carried over into this context is the connection
between spin, labeling the spin networks that are the
kinematical quantum states, and area.

Finally, on the conceptual side, the proposal made here concerning
the interpretation of quantum theory has certain implications that
need to be investigated.  As presented here, the quantum states
are taken to describe information held by, or available to,
certain observers.   This certainly leaves open the possibility that
the quantum theory might turn out to be the statistical mechanics of
a more fundamental theory such as a non-local hidden
variables theory\cite{me-hidden}.  For reasons that have been often
discussed, any departure from the physical postulates of the quantum
theory is likely to be cosmological in
origin\cite{forabner,lotc,me-hidden,spacetime}.
It may thus be hoped that, even if quantum mechanics, expressed in terms
of linear state spaces and operators, is not a fundamental theory, a
formulation of quantum theory that sensibly extends to a theory of
cosmology may be a stepping stone to a deeper theory.

\section*{ACKNOWLEDGEMENTS}

This work grew out of many years of collaboration and
conversations with Louis Crane and Carlo Rovelli.  I am grateful to
Louis Crane for insisting on the usefullness of the categorical
point of view and
to Carlo Rovelli also for the suggestion that quantum gravity might
be formulated axiomatically.  I must thank also David Alpert, Abhay
Ashtekar,
John Baez, Julian Barbour, Steve Carlip
Roumen Borissov, Mauro Carfora,  Ted Jacobson,
Louis Kauffman,
Seth Major, Roger Penrose, Jorge Pullin, Michael Reisenberger,
Simon Saunders,
Chopin Soo, Leonard Susskind and
Gerard 't Hooft for conversations and
suggestions.   Parts of this work were carried out during
visits to the Institute for Advanced Study in Princeton and SISSA
in Trieste, which allowed good conditions for working.  A visit
to Lumine, for which I have to thank Daniel Kastler,
made possible very stimulating conversations with Louis Crane and
Carlo Rovelli.   The opportunity to present
this proposal to the conference in Bielefeld on ``Quantum theory
without observers" was also most helpful.
This work
has been supported in part by the NSF under grant  PHY-93-96246
to Pennsylvania State University.

\end{document}